\documentclass[reprint,aps,pra,showpacs,superscriptaddress]{revtex4-1} 
\usepackage{amsmath,amssymb}
\usepackage{mathrsfs}
\usepackage{bbm}
\usepackage[dvipdfmx]{graphicx, color}
\usepackage{dcolumn}
\usepackage{bm}
\usepackage{enumerate}
\usepackage{theorem}
\newtheorem{theorem}{Theorem}
\newtheorem{lemma}{Lemma}
{}

\def\argmin{\mathop{\mathrm{argmin}}}

\def\Trace{\mathop{\mathrm{Tr}}}

\newcommand{\Probability}{\mathbbm{P}}  
\newcommand{\Expectation}{\mathbbm{E}}  
\newcommand{\Variance}{\mathbbm{V}}

\begin{document}
\title{Precision-guaranteed quantum metrology}



\author{Takanori Sugiyama}
\email{sugiyama@itp.phys.ethz.ch, sugiyama@sigmath.es.osaka-u.ac.jp}
\affiliation{Institute for Theoretical Physics, ETH Zurich, Wolfgang-Pauli-Strasse 27, CH-8093 Zurich, Switzerland} 
\date{\today}
\begin{abstract}
Quantum metrology is a general term for methods to precisely estimate the value of an unknown parameter by actively using quantum resources. 
In particular, some classes of entangled states can be used to significantly suppress the estimation error. 
Here, we derive a formula for rigorously evaluating an upper bound for the estimation error in a general setting of quantum metrology with arbitrary finite data sets. 
Unlike in the standard approach, where lower bounds for the error are evaluated in an ideal setting with almost infinite data, our method rigorously guarantees the estimation precision in realistic settings with finite data. 
We also prove that our upper bound shows the Heisenberg limit scaling whenever the linearized uncertainty, which is a popular benchmark in the standard approach, shows it. 
As an example, we apply our result to a Ramsey interferometer, and numerically show that the upper bound can exhibit the quantum enhancement of precision for finite data.
\end{abstract}
\maketitle

\section{Introduction}\label{Sec:Introduction}
   
   %
   %
   High-precision measurement is one of the most important techniques for developing science and technology.
   Quantum metrology is a general term for methods to precisely estimate the value of an unknown parameter by actively using quantum resources like entanglement and squeezing \cite{Giovannetti2004Science, Giovannetti2011NaturePhoto, Demkowicz2014arXiv}. 
   For example, when we use a separable state on an $N$-partite system in a Ramsey interferometer, an estimation error of phase, $\delta \phi$, scales as $O(1 / \sqrt{N})$ (the standard quantum limit, SQL). 
   On the other hand, when we use an entangled state like a Greenberger-Horne-Zeilinger (GHZ) state with the same number of particles, $\delta \phi$ scales as $O(1 / N)$ (the Heisenberg limit, HL).
   Such quantum enhancement of precision has been experimentally achieved in several quantum systems like quantum optics \cite{Rarity1990PRL}, ions \cite{Meyer2001PRL}, and atoms \cite{Widera2004PRL}.

   %
   %
   One of the main goal of quantum metrology theory is to derive a fundamental lower bound on the estimation error.
   So far many different benchmarks for the estimation error have been proposed and analyzed \cite{Demkowicz2014arXiv}. 
   The most popular benchmark is the root mean squared error (RMSE), and there are two standard approaches for analyzing the RMSE.
   One is to apply a linear approximation of an estimation method to the RMSE.
   The approximated RMSE is called a linearized uncertainty (LU).  
   The other is to analyze the classical and quantum Cram\'{e}r-Rao bounds (CRBs), which are lower bounds on the RMSE for a class of estimation methods.
   LU and CRBs show the SQL scaling for separable states and the HL scaling for some entangled states.
   
   %
   %
   From a theoretical viewpoint, LU and CRBs both are interesting and important quantities.
   From an experimental viewpoint, however, there are two problems with the use of these quantities.
   First problem is about their $\phi$-dependency.
   LU and CRB are functions of the parameter to be estimated.
   The true value of the parameter is unknown in experiments, which is the reason why we try to estimate it.
   This means that we cannot know the exact values of LU and CRB in experiments, although it is possible to estimate their values from experimental data.
   Second problem is about their invalidity for finite data.
   In experiments an amount of available data is finite.  
   A linear approximation is used in the derivation of the LU, while many estimation methods used in quantum metrology like a maximum-likelihood estimator are nonlinear functions of data, and the nonlinearity is not negligible for finite data.
   CRBs are lower bounds of the RMSE, and they are not attainable when the amount of data is finite \footnote{There are two types of the CRBs. One is for finite data, and the other is for infinite data. The sufficient and necessary condition for the attainability of CRBs for finite data that the probability distribution of measurement outcomes is an affine function of the parameter $\phi$ \cite{AmariNagaoka_text2000}. In quantum metrology the probability distribution is not affine of $\phi$, and the CRBs are not attainable for finite data. It becomes attainable at infinite data.}. 
   Unattainable lower bounds on an estimation error cannot be used to guarantee an estimation precision.
   Because the final goal of quantum metrology experiments is a highly precise estimation of an unknown parameter, it is best to {\it rigorously} guarantee an estimation precision, if possible. 
   In order to do that, we need an upper bound on an estimation error satisfying two conditions: (1) be independent of the unknown parameter $\phi$, and (2) be valid for finite data.

   In this paper, we derive an upper bound satisfying these two conditions for a general setting in quantum metrology.
   In Sec. \ref{Sec:Preliminaries}, we explain the setting, notation, and our approach.
   In Sec. \ref{Sec:Results}, we introduce an estimation method called a least squares estimator and give a theorem about the estimator.    
   The upper bound shown in the theorem makes it possible to rigorously guarantee the estimation precision in experiments with finite data, which is not possible by the standard approach of quantum metrology theory. 
   We sketch the proof, and the details are given in Appendix \ref{sec:ProofTheorem1}.
   In Sec. \ref{Sec:Analysis}, we prove that the upper bound shows the scaling same as the LU, which means that the upper bound shows the HL scaling whenever the LU shows it. 
  As an example, we apply our method to a Ramsey interferometer with $N$ atoms, and perform Monte Carlo simulations for $N=1\sim 100$.
  The numerical results indicate that the upper bound can exhibit the quantum enhancement of precision for finite data.  
  In Sec. \ref{Sec:Discussion}, we discuss how to treat known and unknown systematic errors in our approach.
  We summarize this paper In Sec. \ref{Sec:Summary}.

\section{Preliminaries}\label{Sec:Preliminaries}

   \subsection{Procedures and assumptions}
\begin{figure}[bt]
         \centering
         \includegraphics[width=\linewidth]{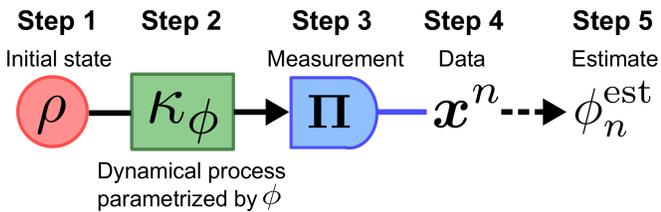}%
         \caption{A procedure for quantum metrology: 
         first, we prepare a known quantum state $\rho$ on a probe system. 
         Second, the state undergoes a dynamical process with an unknown parameter $\phi$.
         Third, we perform a known measurement on the probe system and obtain an outcome.
         Fourth, we repeat the steps described above $n$ times. Then we obtain data consisting of $n$ outcomes.
         Finally, we calculate an estimate of $\phi$ from the data with an estimation method.
         \label{Fig:1}}
      \end{figure}

  We consider the following procedure of quantum metrology (Fig. \ref{Fig:1}):
  \begin{enumerate}[Step. 1]
   \item Prepare a known quantum state $\rho$ on a probe system.
   \item The state undergoes a dynamical process $\kappa_{\phi}$ with an unknown parameter $\phi$. Our aim is to estimate $\phi \in \Phi := [\phi_{\min}, \phi_{\max}]$, where $\phi_{\min}$ and $\phi_{\max}$ are upper and lower values of possible $\phi$ and are assumed to be known.
   \item After the dynamical process, the state changes to a state $\rho_{\phi} = \kappa_{\phi} (\rho)$, which depends on $\phi$.
         We perform a known measurement on the state and obtain a measurement outcome.
   \item Repeat steps 1 to 3 a number $n$ of times \footnote{Note that $n$ and $N$ are different. $N$ is the number of particles in a probe system used for each measurement trial, and $n$ is the number of measurement trials.}.
         Then we have data consisting of $n$ outcomes, $\bm{x}^{n} = \{ x_{1}, \ldots, x_{n} \}$. 
   \item Calculate an estimate of the parameter, $\phi^{\mathrm{est}}_{n} (\bm{x}^{n})$, from the data by a data processing method $\phi^{\mathrm{est}}$. This function from data to a real value is called an estimator.
   \end{enumerate}

   The measurement performed in Step. 3 is described by a positive operator-valued measure (POVM) $\bm{\Pi} = \{ \Pi_{x} \}_{x \in \mathcal{X}}$, which is not necessarily a projective measurement. 
   We assume that the measurement outcomes are bounded, i.e., $- \infty < a \le x \le b <+\infty$.
   This assumption is valid in practice, because there are technical cutoffs on observable values of measurement outcomes in any experiment.
   The probability distribution is given by $p(x | \rho_{\phi}, \bm{\Pi}) = \Trace [ \rho_{\phi} \Pi_{x}]$.
   Let $\Expectation [ \bm{\Pi} | \rho_{\phi}] $ and $\Variance [\bm{\Pi} | \rho_{\phi}]$ denote the expectation and variance of measurement outcome of $\bm{\Pi}$, respectively.
   The expectation is a function of $\phi$, and let $f$ denote the function, i.e., $f(\phi) := \Expectation \left[ \bm{\Pi} | \rho_{\phi}\right]$.
   We assume that $f$ is injective and that the derivative $\frac{df}{d\phi}$ does not take zero for any $\phi \in \Phi$.
   Let $\mathcal{R}_{f}$ denote the range of $f$, i.e., $\mathcal{R}_{f} := \{ f(\phi) | \phi \in \Phi\}$.
   Then $g := f^{-1}$ exists for $\mathcal{R}_{f}$, and we have $\phi = g \left( \Expectation \left[ \bm{\Pi} | \rho_{\phi}\right] \right)$.
   So, if we know the value of the expectation, we can calculate the value of $\phi$.

   \subsection{Estimators}
   For given data $\bm{x}^{n}$, we define the sample mean $S_{n} := \frac{1}{n} \sum_{i=1}^{n} x_{i}$, where $x_{i}$ is the $i$-th outcome in the $n$ outcomes.
   The sample mean converges to the expectation $\Expectation [\bm{\Pi} | \rho_{\phi}]$ in the limit of $n$ going to infinity (the law of large numbers).
   It might seem natural to consider a direct inversion estimator, $\phi^{\mathrm{DI}}_{n} (\bm{x}^{n}) := g (S_{n})$.
   In general, however, the direct inversion estimator does not work well, because the sample mean is a random variable and can be out of $\mathcal{R}_{f}$, which is caused by a statistical fluctuation originated from the finiteness of $n$. 
   The inverse function $g=f^{-1}$ may not exist outside $\mathcal{R}_{f}$, and we may not be able to calculate $\phi^{\mathrm{DI}}_{n} (\bm{x}^{n})$ there.
   Even if $g$ exists, $\phi^{\mathrm{DI}}_{n} (\bm{x}^{n})$ can be out of $\Phi$.
   
   One solution to avoid this problem of $\phi^{\mathrm{DI}}$ mentioned above is a maximum-likelihood estimator (MLE), $\phi^{\mathrm{ML}}_{n} (\bm{x}^{n}) := \argmin_{\phi^{\prime} \in \Phi} \prod_{i=1}^{n} p(x_{i} | \rho_{\phi^{\prime}}, \bm{\Pi})$.
   Unlike $\phi^{\mathrm{DI}}_{n}$, an estimate of the MLE always exists and takes a value in $\Phi$. 
   The MLE has good statistical properties, for example, it attains the Cram\'{e}r-Rao bound in the limit of $n$ going to infinity \cite{RaoText1973}.
   The asymptotic ($n \sim \infty$) behavior of the MLE is well known in classical statistics \cite{Bahadur1960}, but a rigorous analysis for finite $n$ is an open problem.
   Instead of the MLE, we consider a different estimator relatively easier to be analyzed.

   \subsection{Standard benchmarks}\label{subsec:StandardBenchmarks}
   
   Let us choose an estimator $\phi^{\mathrm{est}}$.
   The estimates $\phi^{\mathrm{est}}_{n}(\bm{x}^{n})$ depend on data and probabilistically fluctuate.
   This means that we can observe estimates deviated from the true parameter $\phi$.  
   This difference is called an estimation error of the estimator.
   To evaluate an estimation error is an important topic in quantum metrology.
   
   As explained in Sec \ref{Sec:Introduction}, the most popular benchmark is the root mean squared error (RMSE), which is defined as 
   \begin{eqnarray}
      (\delta \phi)_{\mathrm{RMSE}} :=\sqrt{\Expectation \left[ \left| \phi^{\mathrm{est}}_{n} - \phi \right|^{2}\right]}.
   \end{eqnarray} 
   Generally speaking, a direct analysis of the RMSE itself is difficult, since the RMSE is a function of the dynamics $\kappa_{\phi}$ and our choice of the initial state $\rho$, measurement $\bm{\Pi}$, and estimator $\phi^{\mathrm{est}}$.
   There are two approaches to reduce the degree of this difficulty for analyzing the RMSE. 
   
    The first approach is to approximate the RMSE for the DI estimator.
    When the number of repetition $n$ is sufficiently large, we can approximate the RMSE of the DI estimator as follows: 
    \begin{eqnarray}
       (\delta \phi)_{\mathrm{RMSE}} (\rho , \kappa_{\phi}, \bm{\Pi}, \phi^{\mathrm{est}} = \phi^{\mathrm{DI}}, n)      
       \approx (\delta \phi)_{\mathrm{LU}}:=\frac{B_{\mathrm{LU}}}{\sqrt{n}}, \label{eq:approx_LU}
    \end{eqnarray}
    where
    \begin{eqnarray}
       B_{\mathrm{LU}} := \frac{\sqrt{\Variance [\bm{\Pi} | \rho_{\phi}]}}{\left| \frac{d}{d\phi} \Expectation [\bm{\Pi} | \rho_{\phi}] \right|}.       
    \end{eqnarray}
    In this approximation, the nonlinearity of the DI estimator is ignored.
    In other words, the R.H.S. of the Eq. (\ref{eq:approx_LU}) is the RMSE of the linearized DI estimator, and it is called the linearized uncertainty (LU).
    The details of this approximation are given in Appendix \ref{sec:LU}.
    An advantage of the LU is that the analysis is easy, because it consists of the variance and derivative of expectation with respect to a single outcome.
    Estimates of $B_{\mathrm{LU}}$ such as 
    \begin{eqnarray}
        \left. \frac{\sqrt{\Variance [\bm{\Pi} | \rho_{\phi}]}}{\left| \frac{d}{d\phi} \Expectation [\bm{\Pi} |. \rho_{\phi}] \right|} \right|_{\phi = \phi^{\mathrm{DI}}_{n}(\bm{x}^{n})} \label{eq:Estimate_LU}
    \end{eqnarray}
    are calculated in some experiments \cite{Taylor2008_NaturePhysics,Napolitano2010Nature,Xiang2011NaturePhotonics}.
    
    The second approach is to analyze an asymptotic lower bound of the RMSE, which is independent of our choice of estimator.
    Let us introduce two classes of estimators in statistical estimation theory.
    When the expectation of an estimator with respect to $n$ equals to the true parameter, the estimator is called unbiased for $n$.
    When the derivative of the expectation converges to one in the limit of $n$ going to the infinity, the estimator is called asymptotically unbiased.
    For any unbiased estimator with respect to $n$, the following inequality holds under certain normal conditions \cite{RaoText1973,Helstrom76,Holevo82}:    
    \begin{eqnarray}
       (\delta \phi)_{\mathrm{RMSE}} (\rho , \kappa_{\phi}, \bm{\Pi}, \phi^{\mathrm{est}}, n)
       &\ge& 1 / \sqrt{n \cdot F_{\mathrm{C}} (\phi, \rho , \bm{\Pi})} \label{eq:CCRB} \\
       &\ge& 1 / \sqrt{n \cdot F_{\mathrm{Q}} (\phi, \rho)}, \label{eq:QCRB}
    \end{eqnarray}    
    where $F_{\mathrm{C}}$ and $F_{\mathrm{Q}}$ are quantities called the classical and quantum Fisher information, respectively.
    The Eqs. (\ref{eq:CCRB}) and (\ref{eq:QCRB}) are called the classical and quantum Cram\'{e}r-Rao inequality for finite $n$, respectively.  
    Most of estimators in quantum metrology, which include DI an MLE, are biased for finite $n$, which is originated from the nonlinear parametrization of probability distributions.
    This means that the Cram\'{e}r-Rao inequalities for finite $n$ is not applicable for quantum metrology.
    However, most of ``natural" estimators in quantum metrology, which includes DI and MLE again, are asymptotically unbiased.
    For any asymptotically unbiased estimator, the following inequality holds under certain normal conditions \cite{RaoText1973,Helstrom76,Holevo82}:
    \begin{eqnarray}
       \lim_{n\to \infty} \sqrt{n}\cdot (\delta \phi)_{\mathrm{RMSE}} &(\rho& , \kappa_{\phi}, \bm{\Pi}, \phi^{\mathrm{est}}, n) \notag \\   
       &\ge& 1 / \sqrt{F_{\mathrm{C}} (\phi, \rho , \bm{\Pi})} \label{eq:AsymptoticCCRB}\\
       &\ge& 1 / \sqrt{F_{\mathrm{Q}} (\phi, \rho)}. \label{eq:AsymptoticQCRB}
    \end{eqnarray}
    where Eqs. (\ref{eq:AsymptoticCCRB}) and (\ref{eq:AsymptoticQCRB}) are called classical and quantum Cram\'{e}r-Rao inequalities for asymptotic $n$, respectively.
    The MLE attains the classical Cramer-Rao bound (CRB) for asymptotic $n$ \cite{RaoText1973}.
    An advantage of the CRBs is their generality for our choice of estimator and measurement.
    The classical CRB is independent of estimators, and the quantum CRB is independent of measurement.
    So, classical and quantum CRBs are used for evaluating ultimate performances of a combination of $\rho$ and $\bm{\Pi}$ or $\rho$, respectively.

     From a theoretical viewpoint, LU and CRBs are interesting and important quantities.
     However, they are not suitable for rigorously evaluating an estimation error in experiments with finite $n$ because of the following two reasons.
     (1) LU and CRBs are functions of the unknown parameter $\phi$, and we cannot know their exact values in experiments. 
     Of course, we can estimate their values by calculating quantities like Eq. (\ref{eq:Estimate_LU}), but the calculated values are estimates that can be different from the exact value.
     (2) LU and CRBs are not valid for finite $n$ as explained in this subsection. 
     When $n$ is sufficiently large, we may be able to validate the use of them, but it is unclear which $n$ can be interpreted as sufficiently large.
     In order to rigorously evaluate an estimation error for finite $n$, we need another benchmark.

   \subsection{Confidence intervals}\label{subsec:ConfidenceIntervals}  

    Roughly speaking, confidence intervals are intervals including the true parameter with high probability.
    We propose the size of an confidence interval as a new benchmark in quantum metrology.   
    When an interval $I$ is a function of data and is independent of $\phi$, the function is called an interval estimator.
    We would like to find an interval estimator such that the interval estimates $I(\bm{x}^{n})$ include $\phi$ with high probability.
    When $I(\bm{x}^{n}) \ni \phi$ holds with probability at least $1-\epsilon$ for any $\phi \in \Phi$, the interval estimator is called a confidence interval with ($1-\epsilon$)-confidence level.
    For example, $I=\Phi$ is a confidence interval with $1$-confidence level. 
    This example is trivial and useless.
    We need a nontrivial and useful confidence interval.    
    The following two properties are required for a ``nontrivial" and ``useful" confidence interval in quantum metrology experiments.
    \begin{enumerate}[Property. 1]
       \item Its size converges to zero in the limit of $n$ going to infinity.
       \item Its size can show a quantum enhancement when we use a quantum resource in quantum metrology experiments.  
    \end{enumerate}
    In this paper, we propose a new confidence interval and prove that it has two properties mentioned above.

    Before moving on to our results, let us note their difference from known results.
   Confidence interval and confidence level are well known concepts in classical statistics, and there are many statistical techniques to calculate them for finite data \cite{Lehmann_textTesting2005}.
   Most of these techniques are, however, based on the normal distribution approximation (NDA), and a confidence interval calculated with the NDA is called an approximate confidence interval.
   The NDA is valid when the number of measurement trials $n$ is sufficiently large (the central limit theorem), but again, it is not clear which $n$ can be considered as sufficiently large.
   Therefore, it is not rigorous to apply approximate confidence intervals for finite data in experiments.
   In contrast to an approximate confidence interval, a confidence interval calculated without any assumption on probability distribution is called an exact confidence interval.
   Our new confidence interval is an exact confidence interval, and to the best of our knowledge, it is the first exact confidence interval for quantum metrology.  
   
   An exact confidence region, which is a generalization of confidence interval to higher dimensional spaces, for quantum tomography was proposed in \cite{Sugiyama2013_PRL}.
   The estimation object in quantum tomography is quantum state, process, or measurement, which includes multi parameters.
   Some readers might think that the result in \cite{Sugiyama2013_PRL} would be applicable for quantum metrology because quantum metrology is an estimation problem of quantum process with single parameter, but this is not correct.
   Quantum process tomography and quantum metrology are different problems from statistical viewpoints, and the result for quantum tomography obtained in \cite{Sugiyama2013_PRL} is not applicable for quantum metrology.
   The main difference is from the difference of their parametrization of probability distribution.   
   In quantum tomography, the probability distribution of measurement outcome can be linearly parametrized by the estimation object.
   In quantum metrology, on the other hands, the estimation object is single parameter, but the parametrization of probability distribution is nonlinear.   
   In general, statistical estimation problems with linearly parametrized and nonlinearly parametrized probability distributions have different statistical properties.
   For example, the classical CRB for finite data is attainable in the linear case, but it is not attainable in the nonlinear case \cite{AmariNagaoka_text2000}. 
   Actually, in quantum tomography, there exists an unbiased estimator that attains the equality of the classical Cramer-Rao inequality for any finite data, but in quantum metrology, there do not exist any unbiased estimators that attain the equality for finite data.
   This is caused from the difference of their parametrizations.  
   Therefore quantum process tomography and quantum metrology are different problems in statistical estimation.   
   Additionally, the linearity of the parametrization in quantum tomography is used in the derivation of the exact confidence region in \cite{Sugiyama2013_PRL}, and the result is not applicable for quantum metrology.   
   In order to derive an exact confidence interval that reflects quantumness of resources in quantum metrology, we need new mathematical techniques.

\section{Results}\label{Sec:Results}
We consider the least squares (LS) estimator,
\begin{eqnarray}
   \phi^{\mathrm{LS}}_{n} (\bm{x}^{n}) := \argmin_{\phi^{\prime} \in \Phi} \left| S_{n} - f(\phi^{\prime}) \right|.
\end{eqnarray}
Same as the MLE, the LS estimates $\phi^{\mathrm{LS}}_{n} (\bm{x}^{n})$ always exist and take values in $\Phi$.
Let us define
\begin{eqnarray}
   S_{n}^{\mathrm{LS}} &:=& \argmin_{r^{\prime} \in \mathcal{R}_{f}} \left| r^{\prime} - S_{n} \right|, \\
   V_{\max} &:=& (b-a)^{2}/4, \\   
   V_{n} &:=& \frac{1}{n-1}\sum_{i=1}^{n} \left( x_{i} - \frac{1}{n}\sum_{j=1}^{n}x_{j} \right)^{2},\\
   L &:=& \max_{\phi^{\prime} \in \Phi} \left| \left\{ \frac{df}{d\phi} (\phi^{\prime}) \right\}^{\! -3}\! \! \! \cdot \frac{d^{2}f}{d\phi^{2}} (\phi^{\prime}) \right|. 
\end{eqnarray}
Note that $V_{n}$ is the unbiased sample variance of the data and satisfies $\Expectation [V_{n}] = \Variance \left[ \bm{\Pi} | \rho_{\phi} \right]$.   
Using the quantities introduced above, we define three functions of data $\bm{x}^{n}$ and an user-specified constant $\epsilon$.\begin{eqnarray}
   \delta_{1}(\bm{x}^{n}, \epsilon) 
   &:=& \frac{1}{\left| \frac{d f}{d\phi} (\phi^{\mathrm{LS}}_{n})\right|}  \sqrt{\frac{2}{n} V_{\max} \ln \frac{2}{\epsilon}} 
            + \frac{L}{n}V_{\max} \ln \frac{2}{\epsilon}, \\
   \delta_{2} (\bm{x}^{n}, \epsilon)
   &:=&  \frac{1}{\left| \frac{d f}{d\phi} (\phi^{\mathrm{LS}}_{n})\right|} 
            \left\{ \sqrt{\frac{2}{n}V_{n} \ln \frac{4}{\epsilon}} + \frac{8(b-a)}{3(n-1)} \ln \frac{4}{\epsilon}\right\} \notag \\
   &\phantom{:=}& + \frac{L}{2} \left\{ \sqrt{\frac{2}{n}V_{n} \ln \frac{4}{\epsilon}} + \frac{8(b-a)}{3(n-1)} \ln \frac{4}{\epsilon}\right\}^{2}, \\
   \delta (\bm{x}^{n}, \epsilon) 
   &:=& \left\{ 
   \begin{array}{ll}
      \delta_{1}(\bm{x}^{n}, \epsilon) & \mbox{if}\ n = 1 \\   
      \min \left\{ \delta_{1}(\bm{x}^{n}, \epsilon), \delta_{2} (\bm{x}^{n}, \epsilon) \right\} & \mbox{if}\ n \ge 2
   \end{array}
   \right.\label{def:delta}
\end{eqnarray}
%
%
The following theorem guarantees that the deviation of the LS estimates from the true parameter is upper bounded by $\delta$ with high probability.
\begin{theorem}\label{Theorem:ExactConfidenceInterval}
   For any number of measurement trials $n \ge 1$,  user-specified constant $\epsilon \in (0,1)$ , and true parameter $\phi \in \Phi$,
   \begin{eqnarray} 
      | \phi^{\mathrm{LS}}_{n} (\bm{x}^{n}) - \phi | \le \delta (\bm{x}^{n}, \epsilon)
   \end{eqnarray}
   holds with probability at least $1 - \epsilon$.
\end{theorem}
Theorem 1 means that the $\delta$ defined in Eq. (\ref{def:delta}) is a rigorous ``error bar" on LS estimates $\phi^{\mathrm{LS}}_{n}(\bm{x}^{n})$ for arbitrary finite data.
On the other words, $\phi = \phi ^{\mathrm{LS}}_{n}(\bm{x}^{n}) \pm \delta (\bm{x}^{n}, \epsilon)$ holds with high probability $1-\epsilon$, where we choose the value of $\epsilon$ as small as we like.
The $\delta (\bm{x}^{n}, \epsilon)$ becomes larger as we choose smaller $\epsilon$.
This means that, if we require a higher confidence level for a fixed $n$, the ``error bar" becomes larger for safe.
If we want to keep the confidence level, we need to increase the number of measurement.
   
%
%
We sketch the proof of Theorem 1, with the details shown in the Appendix \ref{sec:ProofTheorem1}.
The LS estimator is a nonlinear function of the sample mean, which is the origin of the main difficulty for the analysis.
We use the Taylor expansion up to the second order with the remainder, and reduce the problem to an analysis on the deviation of the sample mean from the true expectation, $| S_{n} - f(\phi)|$.
In the reduction, we use the contractivity of the LS estimator, i.e.,
\begin{eqnarray}
   | S_{n}^{\mathrm{LS}} - f(\phi)| \le | S_{n} - f(\phi) |,\ \forall \bm{x}^{n},\ \phi \in \Phi. \label{eq:NonExpandability}
\end{eqnarray}
The contractivity is one of the two main keys in this proof, and this is the reason why we choose the LS estimator.
After the reduction, we use two inequalities for evaluating $| S_{n} - f(\phi)|$.
One is Hoeffding's inequality \cite{Hoeffding1963}, which is well known in classical statistics. 
The other is the empirical Bernstein inequality \cite{Maurer2009}, which is a new mathematical tool developed for finite data analysis in machine learning.
The empirical Bernstein inequality is the second key in this proof. 
It enables us to show a relation to the linearized uncertainty explained later.
By combining these inequalities, contractivity, and Taylor expansion, we obtain Theorem 1.

%
%
It is important that $\delta (\bm{x}^{n}, \epsilon)$ depends only on data $\bm{x}^{n}$ and user-specified constant $\epsilon$, and that it is independent of the true parameter $\phi$. 
(The probability distribution of $\delta (\bm{x}^{n}, \epsilon)$ depends on $\phi$.)
So, we can calculate $\delta (\bm{x}^{n}, \epsilon)$ without knowing $\phi$.  
Let us introduce a data-dependent interval, 
\begin{eqnarray}
   I_{\epsilon} (\bm{x}^{n}) := \Phi\! \cap\!\! \left[ \phi^{\mathrm{LS}}_{n}(\bm{x}^{n})\! -\! \delta (\bm{x}^{n}\!\!, \epsilon) , \phi^{\mathrm{LS}}_{n}(\bm{x}^{n})\! +\! \delta (\bm{x}^{n}\!\!, \epsilon) \right].
\end{eqnarray}
Theorem 1 guarantees that this interval estimator $I_{\epsilon}$ is an exact confidence interval with ($1-\epsilon$)-confidence level.
For example, when we choose $\epsilon = 0.01$, we obtain a confidence interval $I_{\epsilon = 0.01}(\bm{x}^{n})$ that includes $\phi$ with probability at least $99\%$.
What we do after Step. 4 in quantum metrology experiments is to choose a value of $\epsilon$ as we like and to calculate the LS estimate $\phi^{\mathrm{LS}}_{n}(\bm{x}^{n})$ and $\delta (\bm{x}^{n}, \epsilon)$ from data obtained.
Then we have an estimate of the unknown parameter $\phi$ with a rigorous error bar.


\section{Analysis}\label{Sec:Analysis}
%
The main purpose of this paper is to propose an exact confidence interval satisfying Properties 1 and 2 explained in Sec. \ref{subsec:ConfidenceIntervals}.
By definition of $\delta$ in Eq. (\ref{def:delta}), our new exact confidence interval $I_{\epsilon}$ satisfies the Property 1.
In this section, we theoretically and numerically prove that $I_{\epsilon}$ also satisfies the Property 2. 
In Sec. \ref{subsec:RelationToLU} we show relations to the LU and quantities calculated in experiments. 
Especially the relation to the LU indicates that $\delta$ shows a quantum enhancement for asymptotically large $n$ whenever the LU shows it.
In Sec. \ref{subsec:RamseyInterferometer}, we perform a numerical simulation of a Ramsey interferometer.
The result indicates that, even for finite $n$, $\delta$ can show the quantum enhancement when a quantum resource is used in quantum metrology.

\subsection{Relation to LU}\label{subsec:RelationToLU}
%
First, we explain a relation between $\delta$ and the LU.
By definition, $\delta_{2}$ decreases as $O(1/\sqrt{n})$, and the coefficient of  the dominant term is given by $\sqrt{2 V_{n} \ln \frac{4}{\epsilon}} / \left| \frac{df}{d\phi}(\phi^{\mathrm{LS}}_{n}) \right|$.
This coefficient converges to $B_{\mathrm{LU}}\sqrt{2 \ln \frac{4}{\epsilon}}$ in the limit of $n$ going to infinity because $V_{n}$ and $\phi^{\mathrm{LS}}_{n}$ converge to $\Variance [\bm{\Pi} | \rho_{\phi}]$ and $\phi$, respectively.
So, we would expect that $\delta$ have the scaling same as the LU with respect to $n$ and $N$. 
Actually we can prove the following inequality:
\begin{eqnarray}
   \lim_{n\to\infty} \left\{  \sqrt{n} \cdot \Expectation \left[ \delta (\bm{x}^{n}, \epsilon) \right]  \right\}
   \le B_{\mathrm{LU}} \sqrt{2 \ln \frac{4}{\epsilon}}, \label{eq:RelationToLU}
\end{eqnarray}
where $\Expectation$ denotes the expectation with respect to data $\bm{x}^{n}$.
The proof is shown in the Appendix \ref{sec:proof_RelationToLU}.
The logic mentioned above and Eq. (\ref{eq:RelationToLU}) guarantee that, on average, $\delta ( \bm{x}^{n}, \epsilon)$ scales same as the LU.
The upper bound $\delta$ shows the HL scaling, whenever the LU shows it.  
   This is important especially in noisy cases.
   The quantum enhancement of precision can be suppressed when the dynamical process $\kappa_{\phi}$ is noisy \cite{Huelga1997}, and recently there are many proposals for recovering the quantum enhancement with respect to the LU \cite{Macchiavello2000QCCM2, Preskill2000arXiv, Matsuzaki2011} and CRB \cite{Chin2012, Chaves2013, Dur2014}. 
   Eq. (\ref{eq:RelationToLU}) indicates that the recovery method with respect to LU also works well for $\delta$.

 %
 %
  Next, we explain a relation between $\delta$ and quantities calculated in experiments.
  As explained in Sec. \ref{subsec:StandardBenchmarks}, estimates of $B_{\mathrm{LU}}$, such as Eq. (\ref{eq:Estimate_LU}), have been calculated in some quantum metrology experiments.
  Such a quantity also appears in $\delta$.
  Let define an estimate of $B_{\mathrm{LU}}$ as 
  \begin{eqnarray}
     B_{\mathrm{LU}}^{\mathrm{est}} 
     : = \frac{\sqrt{V_{n}}}{\left| \frac{df}{d\phi}(\phi^{\mathrm{LS}}_{n}) \right|}.     
  \end{eqnarray}  
  Then the dominant part of $\delta_2$ is rewritten as $B_{\mathrm{LU}}^{\mathrm{est}} \sqrt{\frac{2}{n}\ln \frac{4}{\epsilon}}$.   
  This means that $\delta_{2}$ is a sum of an estimate of $B_{\mathrm{LU}}$ with a coefficient originated from the value of the confidence level and the higher order, i.e., $O(1/n)$, terms.
   The coefficient and higher order terms are corrections for guaranteeing the statistical rigorousness.  
   Therefore our result is not what is totally different from the conventional method in experiments, but it is an extention from such a rough method toward rigorously treating finite data.

      \begin{figure*}[t]
         \centering
         \includegraphics[width=0.95\linewidth]{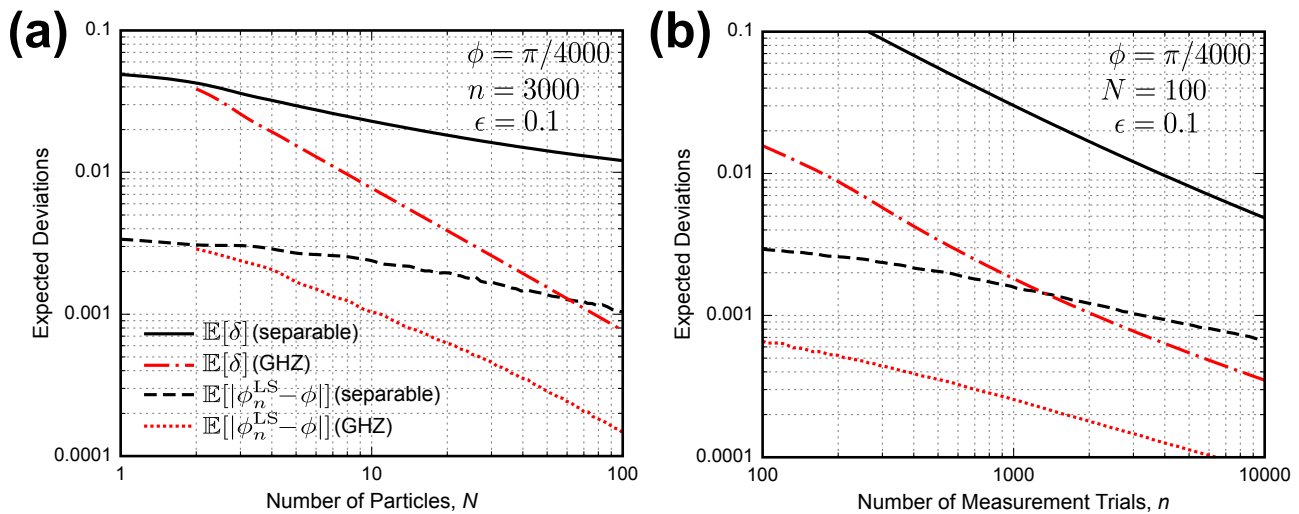}
         \caption{Numerical result on expected deviations ($\Expectation [\delta]$ with $\epsilon = 0.1$, and $\Expectation [|\phi^{\mathrm{LS}}_{n} - \phi|]$) in a Ramsey interferometer using a separable or GHZ initial state of $N$ atoms. 
         Expectations were calculated by a Monte Carlo sampling with $5000$ repetitions. 
         Panel (a) is for their $N$-dependency with $n=3000$, and panel (b) is  for their $n$-dependency with $N=100$. 
         Both panels indicate that $\delta$ for the GHZ state shows the HL scaling up to $N=100$ for finite $n$.
         \label{Fig:4}}
      \end{figure*}     

\subsection{Example: Ramsey interferometer}\label{subsec:RamseyInterferometer}
   We apply our result to a Ramsey interferometer with $N$ atoms.    
   For a separable probe state of $N$ atoms, the LU scales as the SQL scaling, $O(1/\sqrt{N})$.
   On the other hand, for an entangled state like a GHZ state, the LU can scale as the HL scaling, $O(1/N)$.
   We consider two combinations of initial state and measurement.
   One is a combination of a separable state $\left[ \frac{1}{\sqrt{2}}(|e\rangle + |g \rangle )\right]^{\otimes N}$ and the measurement of the total energy, and the other is that of a GHZ state $\frac{1}{\sqrt{2}}(|e \rangle^{\otimes N} + |g \rangle^{\otimes N})$ and the measurement of the parity, where $|e \rangle$ and $|g\rangle$ are excited and ground states of an atom, respectively.   
    In the cases, we have $B_{\mathrm{LU}} = 1/\sqrt{N}$ for the separable state and $B_{\mathrm{LU}}= 1/N$ for the GHZ state.   
   
   We performed Monte Carlo simulations for the cases with $N\! =\! 1\!\sim\! 100$, $n\! =\! 1\! \sim\! 10000$, $\phi_{\min} = 0$, $\phi_{\max} = \pi/400$, $\phi = \pi / 4000$, and $\epsilon = 0.1$ ($90\%$-confidence level).
   The details are given in the Appendix \ref{sec:DetailsOfRamseyInterferometerSimulation}. 
   In order to analyze typical behaviors of $\delta$, we calculated expectations of $\delta$ and compared them to expectations of $|\phi^{\mathrm{LS}}_{n} - \phi |$.
   Fig. \ref{Fig:4} (a) and (b) are the results.   
   In both panels (a) and (b), the vertical axes are for expected deviations. 
   Solid and dashed (black) lines are $\Expectation [\delta (\bm{x}^{n}, \epsilon)]$ and $\Expectation [|\phi^{\mathrm{LS}}_{n}(\bm{x}^{n}) - \phi |]$ for the separable state, respectively.
   Chained and dotted (red) lines are $\Expectation [\delta (\bm{x}^{n}, \epsilon)]$ and $\Expectation [|\phi^{\mathrm{LS}}_{n}(\bm{x}^{n}) - \phi |]$ for the GHZ state, respectively.   
   The expectations were calculated by a Monte Carlo sampling with $5000$ repetitions.
   %
   In panel (a), the horizontal axis is the number of atoms, $N$.
   Plots in the panel express the scaling of the expected deviations with respect to $N$ with a fixed number of measurement trials, $n=3000$.
   The expectations of $\delta$ are larger than those of $|\phi^{\mathrm{LS}}_{n} - \phi |$, which is consistent with Theorem 1.
   Panel (a) also indicates that, up to $N=100$, the expectation of $\delta$ for the GHZ state scales as the HL scaling, although that for the separable state scales as the SQL scaling.
   %
   In panel (b), the horizontal axis is the number of measurement trials, $n$.
   Plots in the panel express the scaling of the expected deviations with respect to $n$ with a fixed number of atoms, $N=100$.
   The expectations of $\delta$ for both states scale as $O(1/\sqrt{n})$, and $\delta$ for the GHZ state is, on average, $10 (= \sqrt{N}$ in the panel) times smaller than $\delta$ for the separable state.
    
    In conclusion of the numerical simulations, the expectations of $\delta$ with 90\%-confidence level are larger than the expectations of the actual deviations for both separable and entangled states, which is consistent with Theorem 1.
   Furthermore, Fig. \ref{Fig:4} indicates that, compared to the separable state, the entangled state gives smaller deviation of estimates and smaller error bar $\delta$.   
   Eq. (\ref{eq:RelationToLU}) guarantees that $\delta$ shows the HL scaling for asymptotically large $n$ whenever the LU shows the scaling, and Fig. \ref{Fig:4} indicates that $\delta$ can also show the quantum enhancement of precision for finite $n$.  
   Note that the Ramsey interferometer is mathematically equivalent to a Mach-Zehnder interferometer \cite{Lee2002JMO}, which means that $\delta$ can show the quantum enhancement of precision in an optical interferometer with a $N00N$ states.

%
%
%

\section{Discussion}\label{Sec:Discussion}

In this section, we discuss how to apply our result to some cases in which unknown systematic errors exist.

\subsection{Partially unknown systematic errors}\label{sec:PartiallyUnkonwnError}

   In Theorem 1, it is assumed that we perfectly know $\rho$, $\bm{\Pi}$, and the functional form of $\kappa_{\phi}$.
   This assumption may not be valid when there exists a systematic error in experiments. 
   In the standard approach of quantum metrology theory, a model for the systematic error is introduced, and it is assumed that the model correctly characterizes the error and that we know the value of a noise parameter in the model \cite{Huelga1997, Macchiavello2000QCCM2, Preskill2000arXiv, Matsuzaki2011, Chin2012, Escher2011, Demkowicz2012, Chaves2013, Dur2014, Arrad2014PRL, Kessler2014PRL}.
   Theorem 1 is applicable for such a perfectly known systematic error.
   Even if the model is correct, however, the value that we think of as the noise parameter may be different from the true value in an experiment.
   Theorem 1 and the standard approach are not directly applicable for such a partially unknown systematic error.
   However, we can obtain an exact confidence interval for quantum metrology with a partially unknown systematic error, by modifying Theorem 1 based on the worst case of the noise parameter.
   
   Here let us consider the case that the noise is partially unknown, i.e., the noise model is correct, but we do not know the value of a noise parameter in the model. 
   Suppose that there is an imperfection of the preparation of initial state and that it is characterized by a noise model with a parameter $\eta_{1}$,
   The time evolution is characterized by the true parameter of interest $\phi$ and noise parameter $\eta_{2}$,
   There is an imperfection in the measurement apparatus, and it is characterized by a noise model with a parameter $\eta_{3}$
   ($\eta_{1}$, $\eta_{2}$, and $\eta_{3}$ can be multi-parameters.)
   Suppose that the noise parameters,$\bm{\eta} := (\eta_{1}, \eta_{2}, \eta_{3})$, are unknown, but that we know a region $E$ including the true noise parameters, i.e., $\bm{\eta} \in E$.
   In this case, the probability distribution of the measurement outcome is given by
   \begin{eqnarray}
      p(x | \phi , \bm{\eta}) = \Trace \left[ \kappa_{\phi , \eta_{2}} (\rho_{\eta_{1}} ) \Pi_{x, \eta_{3}} \right].
   \end{eqnarray}
   We know the function form of the probability distribution, but we do not know the true values of $\phi$ and $\bm{\eta}$.
   Then the functional forms of $f$ and $g$ depends on the values of $\bm{\eta}$, and the value of $\delta$ depends on $\bm{\eta}$ as well.
   To clarify this noise-dependency of $\phi^{\mathrm{LS}}_{n}$ and $\delta$, let us use new notations, $\phi^{\mathrm{LS}}_{n} (\bm{x}^{n}, \bm{\eta})$ and $\delta (\bm{x}^{n}, \epsilon , \bm{\eta})$. 
   
   Let $\bm{\eta}^{\prime} := (\eta_{1}^{\prime}, \eta_{2}^{\prime}, \eta_{3}^{\prime})$ denote the values that we think as the true values of $\bm{\eta}$.
   In general, $\bm{\eta}^{\prime}$ and $\bm{\eta}$ are different. 
   We want to evaluate the difference between $\phi^{\mathrm{LS}}_{n}(\bm{x}^{n}, \bm{\eta}^{\prime})$, which is a LS estimate calculated from data and incorrect noise parameter, and $\phi$.
   We have
   \begin{eqnarray}
      && \left| \phi^{\mathrm{LS}}_{n}(\bm{x}^{n}, \bm{\eta}^{\prime}) - \phi \right| \notag \\
      &\le&  \left| \phi^{\mathrm{LS}}_{n}(\bm{x}^{n}, \bm{\eta}^{\prime}) - \phi^{\mathrm{LS}}_{n}(\bm{x}^{n}, \bm{\eta}) \right|
               + \left| \phi^{\mathrm{LS}}_{n}(\bm{x}^{n}, \bm{\eta}) - \phi \right| \\
      &\le& \left| \phi^{\mathrm{LS}}_{n}(\bm{x}^{n}, \bm{\eta}^{\prime}) - \phi^{\mathrm{LS}}_{n}(\bm{x}^{n}, \bm{\eta}) \right|
               + \delta (\bm{x}^{n}, \epsilon , \bm{\eta}) \label{eq:ConfidenceInterval-SystematicError} \\
      &\le& \max_{\bm{\eta} \in E} 
                \left\{ 
                   \left| \phi^{\mathrm{LS}}_{n}(\bm{x}^{n}, \bm{\eta}^{\prime}) - \phi^{\mathrm{LS}}_{n}(\bm{x}^{n}, \bm{\eta}) \right|
                  + \delta (\bm{x}^{n}, \epsilon , \bm{\eta})                 
                \right\},      
   \end{eqnarray}
   where Eq. (\ref{eq:ConfidenceInterval-SystematicError}) holds with probability at least $1-\epsilon$.
   Let us define
   \begin{eqnarray}
      &&\tilde{\delta} (\bm{x}^{n}, \epsilon, \bm{\eta}^{\prime}) \notag \\
      &&:= \max_{\bm{\eta} \in E} 
                \left\{ 
                   \left| \phi^{\mathrm{LS}}_{n}(\bm{x}^{n}, \bm{\eta}^{\prime}) - \phi^{\mathrm{LS}}_{n}(\bm{x}^{n}, \bm{\eta}) \right|
                  + \delta (\bm{x}^{n}, \epsilon , \bm{\eta})                 
                \right\}. 
                \label{eq:epsilon_PartiallyKnownNoise}  
   \end{eqnarray}
   We obtain the following theorem.
   %
   %
   \begin{lemma}\label{theorem:PartiallyKnownNoise}
      For any number of measurement trials $n \ge 1$, user-specified constant $0 < \epsilon < 1$, unknown true parameter $\phi\in\Phi$, unknown true noise parameters $\bm{\eta} \in E$, and user-specified noise parameters $\bm{\eta}^{\prime}$, 
      \begin{eqnarray}
         \left| \phi^{\mathrm{LS}}_{n}(\bm{x}^{n}, \bm{\eta}^{\prime}) - \phi \right| 
         \le \tilde{\delta}(\bm{x}^{n}, \epsilon, \bm{\eta}^{\prime})          
      \end{eqnarray}
      holds with probability at least $1-\epsilon$.
   \end{lemma}   
   Lemma \ref{theorem:PartiallyKnownNoise} provides an exact confidence interval for quantum metrology with partially unknown noise.
   
   The first term in the R.H.S. of Eq. (\ref{eq:epsilon_PartiallyKnownNoise}) is the effect of the partially unknown noise. 
   This is a systematic error.
   The second term in the R.H.S. of Eq. (\ref{eq:epsilon_PartiallyKnownNoise}) corresponds to the statistical error.
      When a noise is partially unknown and we choose an incorrect value for the noise parameter, any estimator $\phi^{\mathrm{est}}$ cannot converge to the true parameter $\phi$.
      So, when $n$ goes to infinity, $\delta$ converges to $0$ but $\tilde{\delta}$ does not.
      To avoid this problem in the case that the noise is partially unknown, we need to estimate the parameter of interest $\phi$ and noise parameters $\bm{\eta}$ both.    
      This simultaneous estimation of $\phi$ and $\bm{\eta}$ is a theoretically interesting and practically important problem, but it is out of the main topic of this paper.

\subsection{Physical vs statistical models}

   Here, we explain a possible method for treating unknown statistical errors, which is different from the way described in the previous subsection. 
   An experimental setup of quantum metrology is characterized by an initial state $\rho$, a dynamical process $\kappa_{\phi}$, and a measurement $\bm{\Pi}$.
   Let us call a set $(\rho, \kappa_{\phi}, \bm{\Pi})$ a {\it physical model} of the experiment.
   Let us call the function form of the expectation $f$ a {\it statistical model} for the experiment. 
   Recall that the calculation of the LS estimate and $\delta$ requires only $f$.
   So, if we know the statistical model, we can use Theorem 1 even if we do not perfectly know the physical model.

   Our strategy is as follows: 
   \begin{itemize}
      \item Step $1^{\prime}$. We perform a pre-experiment before starting a quantum metrology experiment for an unknown $\phi$. We set a known value of $\phi$ and perform quantum metrology experiments.
      \item Step $2^{\prime}$. We repeat the pre-experiment for many different known values of $\phi$.
      \item Step $3^{\prime}$. We estimate the statistical model $f$ from the data obtained in the pre-experiments.
   \end{itemize}
   If the numbers of known $\phi$s and measurement trials for each known $\phi$ are sufficiently large, we have a precise estimate of the statistical model $f$, and we can use the estimate of $f$ instead of the true $f$ in Theorem 1. 
   The method consisting of Steps from $1^{\prime}$ to $3^{\prime}$ is exactly same as experiments for observing interference fringes in quantum metrology \cite{Leibfried2004,Xiang2011NaturePhotonics,Aiello2012NatureComm}. 
   The precision of estimating $f$ depends on the way of sampling $\phi$s and the choice of estimator for $f$.
   To establish a method for rigorously evaluating a total precision of interference fringe observation and quantum metrology for estimating unknown $\phi$ after the pre-experiments is an open problem, which is important for practical quantum metrology.

\section{Summary}\label{Sec:Summary}
   We considered a general setting of quantum metrology, proposing a least squares estimator and deriving an explicit formula of an exact confidence interval for the estimator with arbitrary finite number of measurement trials.
   The explicit formula makes it possible to calculate a {\it rigorous} error bar, $\delta$, on the least squares estimates in experiments.
   We showed that the error bar $\delta$ scales same as the linearized uncertainty, which is a popular benchmark in the standard approach of quantum metrology, for asymptotically large number $n$ of measurement trials.
   This means that $\delta$ asymptotically shows the Heisenberg limit scaling whenever the linearized uncertainty shows the scaling.
   As an example, we applied our results to a Ramsey interferometer with $N$ atoms and performed Monte Carlo simulations for $N=1\!\sim\! 100$ and $n= 1\!\sim\! 10000$.
   The numerical result indicates that, when a GHZ state is used as an initial state, $\delta$ shows the Heisenberg limit scaling for finite $n$.
   It means that $\delta$ can also exhibit the quantum entrancement of precision for finite $n$.
   To the best of our knowledge, this is the first result that makes it possible to rigorously guarantee an estimation precision in quantum metrology with finite data, and we hope it finds application in the analysis of experimental data.

\section*{Acknowledgements}
    The author would like to thank Patrick Birchall, Hugo Cable, Jonathan Matthews, Javier Sabines, and Peter S. Turner for helpful discussion about possible application of our results to quantum metrology experiments on optical systems and Fuyuhiko Tanaka for useful comments on this paper.  
    The author also thank Martin B. Plenio for drawing his attention to Refs. \cite{Macchiavello2000QCCM2, Preskill2000arXiv}.
    This work was supported by JSPS Postdoctoral Fellowships for Research Abroad (H25-32), the German Science Foundation (grant CH 843/2-1), the Swiss National Science Foundation (grants PP00P2-128455, 20CH21-138799 (CHIST-ERA project CQC)), the Swiss National Center of Competence in Research `Quantum Science and Technology (QSIT)' and the Swiss State Secretariat for Education and Research supporting COST action MP1006.

\appendix

\section*{Appendix}
We explain the details of our results.
In Sec. \ref{sec:NotationAssumption}, we give a summary of assumptions and the derivation of the linearized uncertainty. 
In Sec. \ref{sec:ProofTheorem1}, we give the proof of Theorem 1.
In Sec. \ref{sec:proof_RelationToLU}, we give the proof of Eq. (20).
In Sec. \ref{sec:DetailsOfRamseyInterferometerSimulation}, we explain the details of the Ramsey interferometer and Monte Carlo simulation mentioned in Sec. \ref{subsec:RamseyInterferometer}.

\section{Notations and Assumptions}\label{sec:NotationAssumption}
In this section, for convenience we give a summary of assumptions.
We also explain a relation between the RMSE and LU.

\subsection{List of Assumptions}

   Theorem 1 holds under the following four assumptions.
   \def\theenumi{A\arabic{enumi}}
   \begin{enumerate}
      \item We know $\rho$, $\bm{\Pi}$, and the functional form of $\kappa_{\phi}$, i.e., we know $f$.
      \item The measurement outcomes are bounded, i.e., $-\infty < a \le x \le b < +\infty, \ \forall x \in \mathcal{X}$.
      \item $f$ is injective for $\phi \in \Phi$.
      \item The derivative of $f$ is always non-zero on $\Phi$, i.e., $\left| \frac{df}{d\phi} \right| \neq 0,\ \forall \phi \in \Phi$.
   \end{enumerate}
   \def\theenumi{\arabic{enumi}}   
   Assumption A1 is the standard assumption not only in quantum metrology, but also in statistical parameter estimation.
   Assumption A2 is necessary for the use of Hoeffding's inequality (Lemma 2) and empirical Bernstein inequality (Lemma 3) in the proof of Theorem 1.
   Unbounded outcomes can exist theoretically, but outcomes are always bounded in experiments since there is a technical limit, or cutoff, on an observable range of measurement outcomes.
   So, assumption A2 is natural in experiments.
   Assumption A3 is necessary for the uniqueness of the least squares estimates for any data, and assumption A4 is necessary for avoiding the divergence of $\delta_{1}$ and $\delta_{2}$.
   In Sec. \ref{sec:LU}, we explain that assumptions A3 and A4 are required in the use of the linearized uncertainty, which means that A3 and A4 are implicitly assumed in the standard approach using the LU.

\subsection{Linearized uncertainty and Assumptions}\label{sec:LU}

   We explain a relation between the RMSE and LU (Eq. (\ref{eq:RMSE_LU})), which clarifies the role of assumptions A3 and A4 for the LU.
   The RMSE of an estimator $\phi^{\mathrm{est}}$ is defined by
   \begin{eqnarray}
      (\delta \phi)_{\mathrm{RMSE}}(\phi^{\mathrm{est}}, n | \phi) := \sqrt{\Expectation [ (\phi^{\mathrm{est}}_{n} (\bm{x}^{n}) - \phi)^{2} ]}.
   \end{eqnarray}
   Let us choose the DI estimator $\phi^{\mathrm{DI}}$ as the estimator.
   The DI estimates do not necessarily exist for any data.
   Assumption A3 guarantees the existence of the DI estimates only for $S_{n} \in \mathcal{R}_{f}$. 
   When $S_{n}$ is out of $\mathcal{R}_{f}$, the DI estimate may not exist.
   A3 is a necessary condition for the existence, but it is not a sufficient condition.
   However, let us ignore this fact, i.e., we assume that DI estimates exist for any data.
   By definition, 
   \begin{eqnarray}
      f(\phi^{\mathrm{DI}}_{n}) = S_{n}
   \end{eqnarray} 
   holds.
   We have
   \begin{eqnarray}
      \Variance [\bm{\Pi} | \rho_{\phi}] 
      &=& n \cdot \Expectation [ (S_{n} - \Expectation [\bm{\Pi} | \rho_{\phi}])^{2}] \\
      &=& n \cdot \Expectation [ \{ f(\phi^{\mathrm{DI}}_{n}) - f(\phi) \}^{2}].
   \end{eqnarray}
   We apply the Taylor expansion to $f$,
   \begin{eqnarray}
      f(\phi^{\mathrm{DI}}_{n}) = f(\phi) + \frac{df}{d\phi}(\phi) \cdot (\phi^{\mathrm{DI}}_{n} - \phi ) + O(| \phi^{\mathrm{DI}}_{n} - \phi |^{2}),
   \end{eqnarray}
   and suppose that $n$ is sufficiently large that the nonlinear terms in  the Taylor expansion, $O(| \phi^{\mathrm{DI}}_{n} - \phi |^{2})$, is negligible.
   Then
   \begin{eqnarray}
       \Variance [\bm{\Pi} | \rho_{\phi}] 
       &=& n\cdot \left(\frac{df}{d\phi}  \right)^{2} \Expectation [(\phi^{\mathrm{DI}}_{n} - \phi )^{2}] \\
       &\approx& n\cdot \left(\frac{df}{d\phi}  \right)^{2} (\delta \phi)_{\mathrm{RMSE}}(\phi^{\mathrm{DI}}, n | \phi)^{2} 
   \end{eqnarray}
   holds.
   Since $\frac{df}{d\phi}\neq 0$ holds from assumption A4, we obtain
   \begin{eqnarray}
      (\delta \phi)_{\mathrm{RMSE}}(\phi^{\mathrm{DI}}, n | \phi)
      &\approx& \frac{1}{\sqrt{n}} \frac{\sqrt{\Variance [\bm{\Pi} | \rho_{\phi}]}}{\left| \frac{d}{d\phi} \Expectation [\bm{\Pi} | \rho_{\phi}] \right|} \label{eq:RMSE_LU}\\
      &=& (\delta \phi)_{\mathrm{LU}}.
   \end{eqnarray}
   Eq. (\ref{eq:RMSE_LU}) means that the linearized uncertainty is an approximated RMSE of the DI estimator, which is derived by ignoring     the existence problem of the estimator and the nonlinearity of $f$. 
   This is the reason why we call $(\delta \phi)_{\mathrm{LU}}$ a {\it linearized} uncertainty.

   In the derivation of Eq. (\ref{eq:RMSE_LU}), the following two conditions are required in addition to assumptions A3 and A4.
   \def\theenumi{C\arabic{enumi}}   
   \begin{enumerate}
      \item DI estimates exist for $S_{n} \notin \mathcal{R}_{f}$.
      \item The number of measurement trials $n$ is sufficiently large that the nonlinearity of $f$ around $\phi$ is negligible.  
   \end{enumerate}
   \def\theenumi{\arabic{enumi}}
   In the standard approach using the LU, assumptions A1, A3, A4, and conditions C1 and C2 are implicitly assumed. 
   On the other hand, Theorem 1 does not require C1 and C2.
   Especially the disuse of C2 is important to analyze finite data, because it is unclear which $n$ can be considered as ``sufficiently" large in C2.

\section{Proof of Theorem 1}\label{sec:ProofTheorem1}

In this section we show the proof of Theorem 1.
We derive an upper bound of $\Probability [ \left| \phi^{\mathrm{LS}}_{n} - \phi \right| > \delta ] = 1 - \Probability [ \left| \phi^{\mathrm{LS}}_{n} - \phi \right| \le \delta ]$.
It is difficult to directly analyze this quantity, because $\phi^{\mathrm{LS}}_{n}$ is a nonlinear function of $S_n$ and is a biased estimator.
On the other hand, the following two lemmas hold for $S_n$.
%
%
\begin{lemma}[Hoeffding's inequality \cite{Hoeffding1963}]\label{lemma:Hoeffding}
   Let $X$ be a random variable with $X \in [a, b]$ and $X_{1}, X_{2}, \ldots , X_{n}$ be a sequence of i.i.d. random variables satisfying $X_{i} = X (i = 1, \ldots , n)$, respectively. 
   Then for any $0<\epsilon < 1$ and $n \ge 1$, 
   \begin{eqnarray}
      \Probability \left[ | S_{n} - \Expectation [X] | > \sqrt{\frac{2}{n}V_{\max} \ln \frac{2}{\epsilon}} \right]
      \le \epsilon  \label{eq:Hoeffding_iid}
   \end{eqnarray}
   holds.
\end{lemma}

%
\begin{lemma}[Empirical Bernstein inequality \cite{Maurer2009}]\label{lemma:EmpiricalBernstein}
   Let $X$ be a random variable with $ a \le X \le b$ and $X_{1}, X_{2}, \ldots , X_{n}$ be a sequence of i.i.d. random variables satisfying $X_{i} = X (i = 1, \ldots , n)$, respectively. 
   Then for any $0 < \epsilon < 1$ and $n \ge 2$, 
   \begin{eqnarray}
      \Probability \left[ \left| S_{n} - \Expectation [X] \right| > \sqrt{\frac{2}{n}V_{n}(\bm{X}^{n})\ln \frac{4}{\epsilon}} + \frac{8(b-a)}{3(n-1)} \ln \frac{4}{\epsilon} \right]
      \le \epsilon \label{eq:Bernstein_iid}
   \end{eqnarray}
   holds\footnote{Some coefficients in Eq. (\ref{eq:Bernstein_iid}) are different from the corresponding inequality in \cite{Maurer2009}, because we have a proof of Eq. (\ref{eq:Bernstein_iid}) but we could not prove the original inequality.}, where
   \begin{eqnarray}
      V_{n}(\bm{X}^{n}) := \frac{1}{n-1} \sum_{i=1}^{n} \left( X_{i} - \frac{1}{n} \sum_{j=1}^{n} X_{j} \right)^{2}.
   \end{eqnarray}
\end{lemma}
Note that $V_{\max}$ in Hoeffding's inequality is independent of data, and that $V_{n} (\bm{X}^{n})$ in the empirical Bernstein  inequality is dependent of data.

First, we reduce the analysis of $\left| \phi^{\mathrm{LS}}_{n} (\bm{x}^{n}) - \phi \right| $ to that of $| S_{n} - \Expectation [\bm{\Pi} | \rho_{\phi}] | $.
Let $r$ denote the argument of $g$.
Using the Taylor expansion of $g ( r )$ around $S_{n}^{\mathrm{LS}}$ up to the 2nd order with the remainder in the Lagrange form, we obtain the following inequality.
\begin{eqnarray}
   &&\left| \phi^{\mathrm{LS}}_{n} (\bm{x}^{n}) - \phi \right| \notag \\
   &=& \left| g (S_{n}^{\mathrm{LS}}) - g (r) \right|  \\
   &=& \left| \frac{d g}{dr}(S_{n}^{\mathrm{LS}}) (r - S_{n}^{\mathrm{LS}}) 
           + \frac{1}{2}\frac{d^2 g}{dr^2}(r^{\prime}) (r - S_{n}^{\mathrm{LS}})^{2} \right| \\
   &\le& \left| \frac{d g}{dr}(S_{n}^{\mathrm{LS}}) \right| \cdot \left| r - S_{n}^{\mathrm{LS}} \right| 
           + \frac{1}{2} \left| \frac{d^2 g}{dr^2} (r^{\prime}) \right| \cdot \left| r - S_{n}^{\mathrm{LS}} \right|^{2}, \label{eq:reduction_step1}          
 \end{eqnarray}
 where $r^{\prime}$ is some real number between $r$ and $S_{n}^{\mathrm{LS}}$.
 By combining Eq. (\ref{eq:reduction_step1}) with $\left| \frac{d^2 g}{dr^2} (r^{\prime}) \right| = \left| \left\{ \frac{df}{d\phi} (\phi^{\prime}) \right\}^{\! -3} \cdot \frac{d^{2}f}{d\phi^{2}} (\phi^{\prime}) \right| \le L$ and the contractivity, we obtain  
 \begin{eqnarray}       
   &&\left| \phi^{\mathrm{LS}}_{n}(\bm{x}^{n}) - \phi \right| \notag \\    
   &\le& \left| \frac{d g}{dr}(S_{n}^{\mathrm{LS}}) \right| \cdot \left| S_{n} - r \right| + \frac{1}{2}L \left| S_n - r \right|^{2}.
\end{eqnarray}
Then $\delta < \left| \phi^{\mathrm{LS}}_{n} - \phi \right|$ implies 
\begin{eqnarray}
   \delta < \left| \frac{d g}{dr}(S_{n}^{\mathrm{LS}}) \right| \cdot \left| S_{n} - r \right| + \frac{1}{2}L \left| S_n - r \right|^{2}. \label{eq:epsilon-Sn}
\end{eqnarray}
By solving this quadratic inequality with $\delta > 0$, we can show that Eq. (\ref{eq:epsilon-Sn}) is equivalent to 
\begin{eqnarray}
   \left| S_{n} - r \right| 
   &>&  \frac{1}{L} \left\{ \sqrt{ \left| \frac{dg}{dr} (S_{n}^{\mathrm{LS}}) \right|^2 + 2 \delta L }  -  \left| \frac{dg}{dr} (S_{n}^{\mathrm{LS}}) \right| \right\}. \label{eq:Sn-epsilon}
\end{eqnarray}
By substituting $\delta = \delta_{1}$ and $\delta = \delta_{2}$ into Eq. (\ref{eq:Sn-epsilon}), we obtain
\begin{eqnarray}
 \frac{1}{L} \left\{ \sqrt{ \left| \frac{dg}{dr} (S_{n}^{\mathrm{LS}}) \right|^2 + 2 \delta_{1} L }  -  \left| \frac{dg}{dr} (S_{n}^{\mathrm{LS}}) \right| \right\}  \notag \\ 
   = \sqrt{\frac{2}{n}V_{\max}\ln \frac{2}{\epsilon}}, \\ 
 \frac{1}{L} \left\{ \sqrt{ \left| \frac{dg}{dr} (S_{n}^{\mathrm{LS}}) \right|^2 + 2 \delta_{2} L }  -  \left| \frac{dg}{dr} (S_{n}^{\mathrm{LS}}) \right| \right\} \notag \\  
   = \sqrt{\frac{2}{n}V_{n}(\bm{x}^{n})\ln \frac{4}{\epsilon}} + \frac{8(b-a)}{3(n-1)} \ln \frac{4}{\epsilon} ,&& 
\end{eqnarray}
where we used the equalities $\frac{dg}{dr} =  \left( \frac{df}{d\phi} \right)^{-1}$ from assumption A3 and $\frac{df}{d\phi}(\phi^{\mathrm{LS}}_{n}) \neq 0$ from assumption A4 (note that $\phi^{\mathrm{LS}}_{n} (\bm{x}^{n}) \in \Phi$ holds for any $\bm{x}^{n}$).
From Lemmas \ref{lemma:Hoeffding} and \ref{lemma:EmpiricalBernstein}, we obtain
\begin{eqnarray}
    &&\Probability \left[ \left| \phi^{\mathrm{LS}}_{n} - \phi \right| > \delta \right] \notag \\
   &\le& \Probability \left[ \left| S_{n} - r \right| 
             >  \min \left\{ \sqrt{\frac{2}{n}V_{\max}\ln \frac{2}{\epsilon}}, \right. \right. \notag \\
    && \hspace{15mm} \left. \left. \sqrt{\frac{2}{n}V_{n}(\bm{x}^{n})\ln \frac{4}{\epsilon}} + \frac{8(b-a)}{3(n-1)} \ln \frac{4}{\epsilon} \right\} \right] \\ 
    &\le& \epsilon.
      \label{eq:EP_LS_Sn}
\end{eqnarray}
$\square$

\section{Proof of Eq. (20)}\label{sec:proof_RelationToLU}

   Here we show the proof of Eq. (20):    
   \begin{eqnarray}
      && \lim_{n\to\infty} \left\{ \sqrt{n} \cdot \Expectation \left[ \delta (\bm{x}^{n}, \epsilon) \right] \right\} \notag \\
      &\le& \lim_{n\to \infty} \left\{ \sqrt{n} \cdot \Expectation \left[ \delta_{2} (\bm{x}^{n}, \epsilon) \right] \right\} \\  
      &=& \lim_{n\to\infty} \left\{ \Expectation \left[  \left| \frac{d g}{dr} (S_{n}^{\mathrm{LS}})\right| \sqrt{V_{n}} \right] \right\} \sqrt{2\ln \frac{4}{\epsilon}} \\
      &\le& \lim_{n\to\infty} \left\{ \sqrt{\Expectation \left[ \left| \frac{d g}{dr} (S^{\mathrm{LS}}_{n})\right|^{2} \right]} \sqrt{\Expectation \left[ V_{n}\right]} \right\} \sqrt{2\ln \frac{4}{\epsilon}} \\
      &=& \sqrt{ \lim_{n\to\infty} \Expectation \left[ \left| \frac{d g}{dr} (S^{\mathrm{LS}}_{n})\right|^{2} \right]} 
             \cdot \sqrt{\Variance [\bm{\Pi} | \rho_{\phi}]} \sqrt{2\ln \frac{4}{\epsilon}},
    \end{eqnarray}         
     where we used the Cauchy-Schwarz inequality and the equality $\Expectation [V_{n}] = \Variance [\bm{\Pi}| \rho_{\phi}]$.
     From the Taylor expansion, we have
     \begin{eqnarray}
        \frac{d g}{dr} (S^{\mathrm{LS}}_{n})     
        =  \frac{d g}{dr} ( r ) + O (| S^{\mathrm{LS}}_{n} - r |).         
     \end{eqnarray}
     At the limit of $n$ to infinity, $|S^{\mathrm{LS}}_{n} - r |$ converges to $0$ because of the contractivity, $|S^{\mathrm{LS}}_{n} - r | \le |S_{n} - r|$, and the law of large numbers.    
     Then 
     \begin{eqnarray}
        \lim_{n\to\infty} \Expectation \left[ \left| \frac{d g}{dr} (S^{\mathrm{LS}}_{n})\right|^{2} \right]
        = \left| \frac{d g}{dr} ( r )\right|^2
      \end{eqnarray}             
     holds, and we obtain                 
    \begin{eqnarray}       
    && \lim_{n\to\infty} \left\{ \sqrt{n} \cdot \Expectation \left[ \delta (\bm{x}^{n}, \epsilon) \right] \right\} \notag \\          
      &\le& \left| \frac{d g}{dr} ( r )\right| \sqrt{\Variance [\bm{\Pi} | \rho_{\phi}]} \sqrt{2\ln \frac{4}{\epsilon}}  \\          
      &=& \frac{\sqrt{\Variance [\bm{\Pi} | \rho_{\phi}]}}{\left| \frac{d \Expectation \left[ \bm{\Pi} | \rho_{\phi}\right]}{d\phi} \right|} \sqrt{2\ln \frac{4}{\epsilon}}.      
   \end{eqnarray}
   $\square$

\section{Details of Ramsey interferometer simulation}\label{sec:DetailsOfRamseyInterferometerSimulation}
  
   In this section, we explain the details of a Ramsey interferometer and the Monte Carlo simulation.
   When we use a separable state of $N$ atoms for the initial state, the LU scales as the SQL scaling, $O(1/\sqrt{N})$.
   On the other hand, when we use an entangled state, the LU can scale as the HL scaling, $O(1/N)$.
   The procedure of the Ramsey interferometer is as follows.
   \begin{enumerate}
      \item Prepare an initial state $|\phi \rangle$ of $N$ atoms. Each atom is a two-level system.
      \item Each atom independently undergoes a free evolution, $\exp \left( i\frac{\phi}{2} \sigma_{3} \right)$.
      \item After the evolution, we perform a $\frac{\pi}{2}$-pulse along an axis, $\cos{\phi_{0}}\cdot \sigma_{1} + \sin{\phi_{0}} \cdot\sigma_{2}$, where $\phi_{0}$ is a reference phase to be user-tuned.
      \item Perform a projective measurement of an observable, $A$. 
      \item Repeat 1 to 4 a number $n$ of times.
   \end{enumerate}
   
   We consider the following two combinations of the initial state $|\psi \rangle$ and measured observable $A$.
   \renewcommand{\labelenumi}{(\theenumi)}
   \begin{enumerate}
      \item A product state and energy measurement
      
        Let us choose a product state,
        \begin{eqnarray}
            |\phi \rangle = \left[ \frac{1}{\sqrt{2}} (|e \rangle + |g\rangle) \right]^{\otimes N}
        \end{eqnarray}
        as the initial state, where $|e\rangle $and $|g\rangle$ are the excited and ground states, respectively.
        We observe the total energy, 
        \begin{eqnarray}
           J_{3} := \sum_{j=1}^{N} \sigma_{3}^{(j)},
        \end{eqnarray}   
        where $\sigma_{3}^{(j)} := I^{\otimes (j-1)} \otimes \sigma_{3} \otimes I^{\otimes (N-j)}$.
        The set of possible measurement outcomes is $\mathcal{X} = \{ -N, -(N-1), \ldots, N-1, N  \}$.
        In this combination, the probability distribution is given by
        \begin{eqnarray}
           p(x| \rho_{\phi}, \bm{\Pi}) 
           &=& \frac{N!}{[(N+x)/2]! [(N-x)/2]!} \notag \\
           & & \hspace{-10mm} \times\left\{ \frac{1 + \sin (\phi - \phi_{0})}{2} \right\}^{\frac{N + x}{2}} \left\{ \frac{1 - \sin (\phi - \phi_{0})}{2} \right\}^{\frac{N - x}{2}}, 
        \end{eqnarray}
        and we obtain the following equalities:
        \begin{eqnarray}
          f (\phi) &=& N \sin{(\phi - \phi_{0})}, \label{eq:f_separable} \\
          \frac{df}{d\phi} &=& N \cos{(\phi - \phi_{0})}, \label{eq:dfdphi_separable}\\
          \Variance [\bm{\Pi} |\rho_{\phi}] &=& N \cos^{2}(\phi - \phi_{0}), \label{eq:variance_separable}\\
          V_{\max}&=& N^{2}, \label{eq:Vmax_separable}\\
          B_{\mathrm{LU}} &=& \frac{1}{\sqrt{N}}, \label{eq:BLU_separable}\\
          \left\{ \frac{df}{d\phi} \right\}^{\! -3} \cdot \frac{d^{2}f}{d\phi^{2}}   
          &=& -\frac{1}{N^{2}} \frac{\sin{(\phi - \phi_{0})}}{\cos^{3}(\phi - \phi_{0})}. \label{eq:d2gdr2_separable}
       \end{eqnarray}    
        From Eq. (\ref{eq:f_separable}), $f$ is a periodic function with period $2\pi$.
        In order to satisfy assumptions A3 and A4, the size of $\Phi$ must be at most smaller than $\pi$.       
      
      \item A GHZ state and parity measurement
      
         Let us choose a GHZ state,
         \begin{eqnarray}
            |\psi \rangle = \frac{1}{\sqrt{2}} (| e\rangle^{\otimes N} + |g\rangle^{\otimes N} )
         \end{eqnarray}
         as the initial state.
         We observe the parity, 
         \begin{eqnarray}
            P = (+1)^{N_{g}}(-1)^{N_{e}},
         \end{eqnarray} 
         where $N_{g}$ and $N_{e}$ are the particle number operators for $|g\rangle$ and $|e\rangle$, respectively.
         The set of possible measurement outcomes is $\mathcal{X} = \{-1, +1 \}$.
         In this combination, the probability distribution is given by 
         \begin{eqnarray}
            p(x | \rho_{\phi}, \bm{\Pi}) = \frac{1}{2} \left\{1 + x\cdot \cos{N\left(\phi - \phi_{0} + \frac{\pi}{2}\right)} \right\}.
         \end{eqnarray}
          In this case, we have the following equalities:      
          \begin{eqnarray}
            f (\phi) &=& \cos{N\left(\phi - \phi_{0} + \frac{\pi}{2}\right)}, \label{eq:f_GHZ}\\
            \frac{df}{d\phi} &=& -N \sin{N\left(\phi - \phi_{0} + \frac{\pi}{2}\right)}, \label{eq:dfdphi_GHZ}\\
            \Variance [\bm{\Pi} |\rho_{\phi}] &=& \sin^{2}{N\left(\phi - \phi_{0} + \frac{\pi}{2}\right)}, \label{eq:variance_GHZ}\\
            V_{\max} &=& 1, \label{eq:Vmax_GHZ}\\
            B_{\mathrm{LU}} &=& \frac{1}{N}, \label{eq:BLU_GHZ}\\
            \left\{ \frac{df}{d\phi} \right\}^{\! -3} \cdot \frac{d^{2}f}{d\phi^{2}}   
            &=& \frac{1}{N} \frac{\cos{N\left( \phi - \phi_{0} + \frac{\pi}{2} \right)}}{\sin^{3}N\left(\phi - \phi_{0} + \frac{\pi}{2}\right)}. \label{eq:d2gdr2_GHZ}
         \end{eqnarray}   
         From Eq. (\ref{eq:f_GHZ}), $f$ is a periodic function with period $2\pi / N$.
         In order to satisfy assumptions A3 and A4, the size of $\Phi$ must be at most smaller than $\pi / N$.          
                 
   \end{enumerate}

      \begin{figure*}[t]
         \centering
         \includegraphics[width=0.95\linewidth]{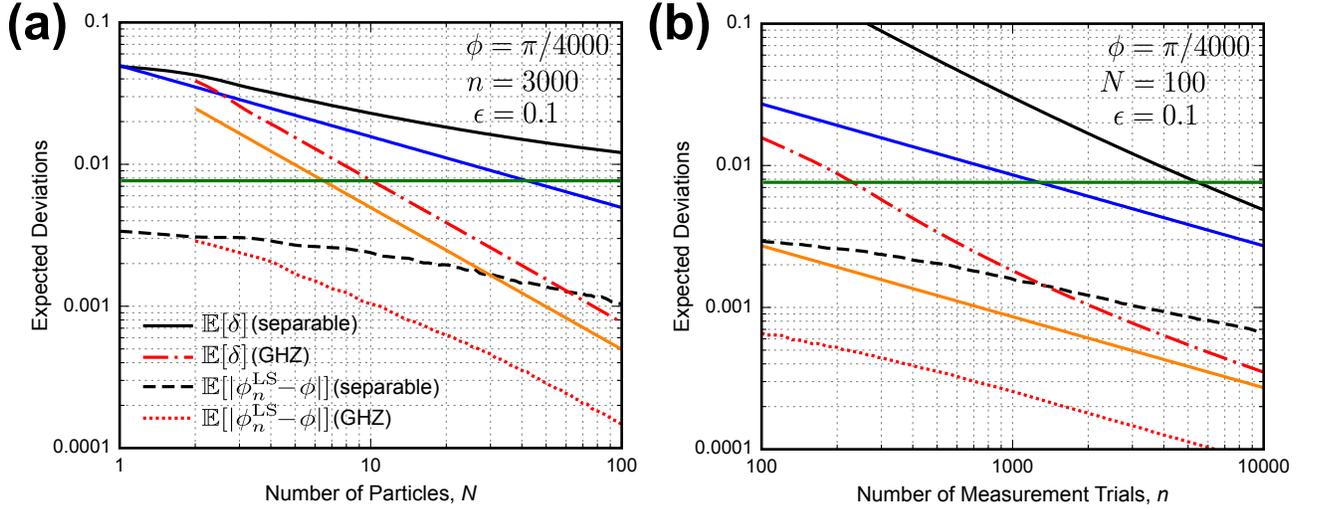}
         \caption{Numerical result on expected deviations ($\Expectation [\delta]$ with $\epsilon = 0.1$, and $\Expectation [|\phi^{\mathrm{LS}}_{n} - \phi|]$) in a Ramsey interferometer using a separable or GHZ initial state of $N$ atoms. 
         Expectations were calculated by a Monte Carlo sampling with $5000$ repetitions. 
         Solid blue and orange lines are $\frac{B_{\mathrm{LU}}}{\sqrt{n}} \sqrt{2\ln \frac{4}{\epsilon}}$ for the separable and GHZ states, respectively.
         Solid green lines are for $\phi_{\max} - \phi_{\min}$, which is a trivial 100\%-confidence interval.
         Panel (a) is for their $N$-dependency with $n=3000$, and panel (b) is  for their $n$-dependency with $N=100$. 
         Both panels indicate that $\delta$ for the GHZ state shows the HL scaling, $O(1/N)$, for finite $n$. 
         \label{Fig:5}}
      \end{figure*}     


   Note that $J_{3}$ and $P$ are commuting, and that these are reduced from the measurement of particle numbers for each energy level in an experiment \cite{Leibfried2004}.
   Mathematically these are different observables, but the same measurement apparatus is used for both of two in the experiment.
   So, in cases (1) and (2), their initial states are different, their observables are different, and their POVMs are same.
   In general, the LU is larger than or equivalent to the CRB, but in cases (1) and (2) their LUs coincide with their CRBs, respectiely.
   This is the reason why we choose different observables.

   We performed a Monte Carlo simulation for cases (1) and (2) with the following parameters: $\phi_{\min} = 0$, $\phi_{\max} = \pi/400$, $\phi = \pi / 4000$, and $\epsilon = 0.1$ (90\%-confidence level).
   The reference phases are chosen as $\phi_{0} = -\pi/8$ for case (1) and $\phi_{0} = \pi/2 - \pi / 10N$.
   We show the result in Fig. \ref{Fig:5}, where $B_{\mathrm{LU}}$ and $\phi_{\max} - \phi_{\min}$ are added to Fig. 2.
   In both panels (a) and (b) of Fig. \ref{Fig:5}, vertical axes are for expected deviations, $\Expectation [\delta (\bm{x}^{n}, \epsilon)]$ and $\Expectation [|\phi^{\mathrm{LS}}_{n}(\bm{x}^{n}) - \phi | ]$.
   Solid and dashed (black) lines are $\Expectation [\delta (\bm{x}^{n}, \epsilon)]$ and $\Expectation [|\phi^{\mathrm{LS}}_{n}(\bm{x}^{n}) - \phi |]$ for the separable state, respectively.
   Chained and dotted (red) lines are $\Expectation [\delta (\bm{x}^{n}, \epsilon)]$ and $\Expectation [|\phi^{\mathrm{LS}}_{n}(\bm{x}^{n}) - \phi |]$ for the GHZ state, respectively.   
   The expectations were calculated by a Monte Carlo sampling with $5000$ repetitions.
   Each horizontal axis in panels (a) and (b) is for the number of atoms $N$ and the number of measurement trials $n$, respectively.
   As explained in Sec. IV B, these panels indicate that $\delta$ for the GHZ state shows the Heisenberg limit scaling, $O(1/N)$ for finite $n$.
   The scalings of $\Expectation [\delta (\bm{x}^{n}, \epsilon)]$ with respect to $N$ and $n$ are independent of $\epsilon$, and the quantum enhancement on $\delta$ appears not only for $\epsilon = 0.1$ but also for other values of $\epsilon$,

   Solid blue and orange lines in Fig. \ref{Fig:5} are $\frac{B_{\mathrm{LU}}}{\sqrt{n}}\sqrt{2 \ln \frac{4}{\epsilon}}$ for cases (1) and (2), respectively.
   Eq. (20) guarantees that $\Expectation [\delta (\bm{x}^{n}, \epsilon)]$ becomes smaller than the new lines in the limit of $n$ going to infinity.
   Panel (b) indicates that two lines for $\Expectation [\delta (\bm{x}^{n}, \epsilon)]$ (solid black and chained red lines) become closer to the blue and orange lines as $n$ becomes larger, respectively.
   However, lines for $\Expectation [\delta (\bm{x}^{n}, \epsilon)]$ are still larger than $\frac{B_{\mathrm{LU}}}{\sqrt{n}}\sqrt{2 \ln \frac{4}{\epsilon}}$ at $n= 10000$.
   This means that $n=10000$ cannot be considered as a ``sufficiently" large number.   
   
   Solid green lines in Fig. \ref{Fig:5} are for $\phi_{\max} - \phi_{\min} = \pi/400 \approx 0.008$.   
   Because $\phi^{\mathrm{LS}}_{n} (\bm{x}^{n})$ is always included in $\Phi$ for any data,  
   \begin{eqnarray}
      \left| \phi^{\mathrm{LS}}_{n} - \phi \right| \le \phi_{\max} - \phi_{\min}
   \end{eqnarray}
   holds with probability $1$.
   So, $\phi_{\max} - \phi_{\min}$ gives a trivial $100\%$-confidence interval.
   In panel (a) of Fig. \ref{Fig:5}, $\Expectation [\delta]$ for the separable state (solid black line) is larger than $\phi_{\max} - \phi_{\min}$ (the solid green line) for all $N$ between $1$ to $100$. 
   This means that, on average, $n=3000$ is not enough for obtaining a nontrivial $90\%$-confidence interval in case (1), while the number is enough in case (2) with $N\ge 10$.
   In panel (b), $\Expectation [\delta]$ for the separable state (solid black line) becomes smaller than $\phi_{\max} - \phi_{\min}$ (solid green line) for $n \ge 6000$.
   If we perform measurement trials more than 6000 times for the separable state, it is expected to obtain a non-trivial $90\%$-confidence interval

      \begin{figure*}[tb]
         \centering
         \includegraphics[width=0.95\linewidth]{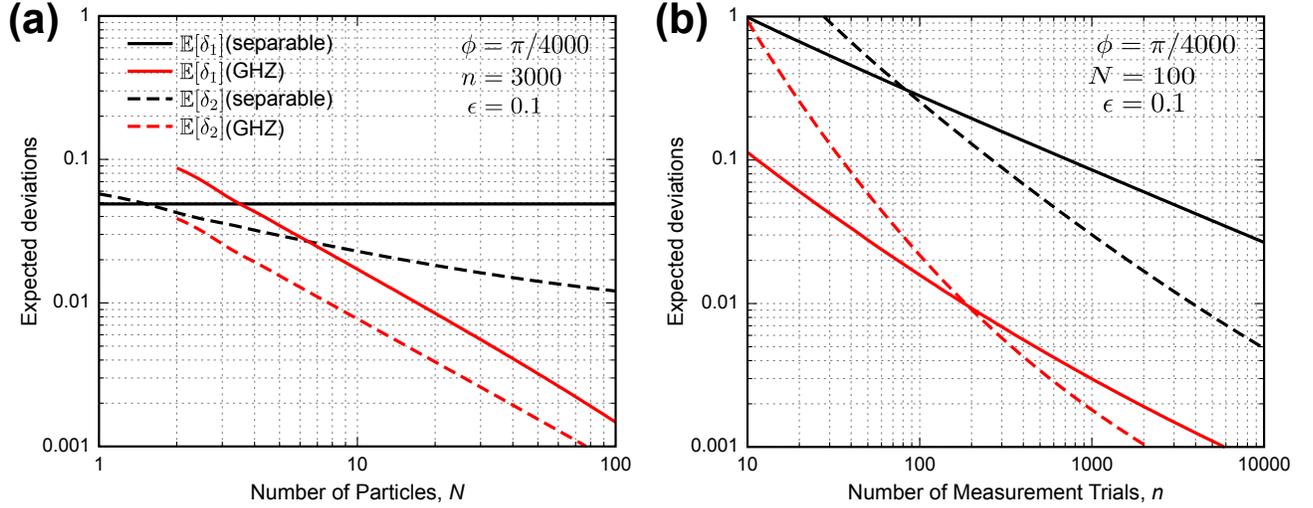}
         \caption{Numerical result on expected deviations ($\Expectation [\delta_{1}]$ and $\Expectation [\delta_{2}]$ with $\epsilon = 0.1$) in a Ramsey interferometer using a separable or GHZ initial state of $N$ atoms. 
         Expectations were calculated by a Monte Carlo sampling with $5000$ repetitions. 
         Panel (a) is for their $N$-dependency with $n=3000$, and panel (b) is  for their $n$-dependency with $N=100$. 
         \label{Fig:6}}
      \end{figure*}     

   Finally we analyze the behaviors of $\delta_{1}$ and $\delta_{2}$.
   In both panels (a) and (b) of Fig. \ref{Fig:6}, vertical axes are for expected deviations, $\Expectation [\delta_{1} (\bm{x}^{n}, \epsilon)]$ and $\Expectation [\delta_{2} (\bm{x}^{n}, \epsilon) ]$.
   Solid and dashed (black) lines are $\Expectation [\delta_{1} (\bm{x}^{n}, \epsilon)]$ and $\Expectation [\delta_{2} (\bm{x}^{n}, \epsilon)]$ for the separable state, respectively.
   Solid and dashed (red) lines are $\Expectation [\delta_{1} (\bm{x}^{n}, \epsilon)]$ and $\Expectation [\delta_{2} (\bm{x}^{n}, \epsilon)]$ for the GHZ state, respectively. 
  In panel (a), the horizontal axis is the number of atoms, $N$.
   Plots in the panel express the scaling of the expected $\delta_{1}$ and $\delta_{2}$ with respect to $N$ with a fixed number of measurement trials, $n=3000$.
   $\Expectation [\delta_{1}]$ for the separable state (solid black) is almost constant, although the other three plots decrease as $N$ becomes large.
   This is caused by the difference between scalings of $V_{\max}$ and $V_{n}$.
   Roughly speaking, $V_{n}$ and $\Variance [\bm{\Pi} | \rho_{\phi}]$ has the same scaling with respect to $N$.
   From Eqs. (\ref{eq:dfdphi_separable}), (\ref{eq:variance_separable}), (\ref{eq:Vmax_separable}), (\ref{eq:d2gdr2_separable}), (D14), (\ref{eq:variance_GHZ}), (\ref{eq:Vmax_GHZ}), and (\ref{eq:d2gdr2_GHZ}), we have the following scalings of $\delta_{1}$ and $\delta_{2}$ with respect to $N$ for fixed $n$s.
   \begin{eqnarray}
      \delta_{1} 
      &=& \left\{
         \begin{array}{ll}
            O(1) & \mbox{(the separable state)} \\
            O(1/N) & \mbox{(the GHZ state)}
         \end{array}
      \right.,\\
      \delta_{2} 
      &=& \left\{
         \begin{array}{ll}
            O(1) & \mbox{(the separable state, small $n$)} \\
            O(1/\sqrt{N}) & \mbox{(the separable state, sufficiently large $n$)} \\
            O(1/N) & \mbox{(the GHZ state)}
         \end{array}
      \right..
   \end{eqnarray}  
   The scaling of dashed black line is between $O(1/\sqrt{N})$ and $O(1)$, which means that $n=3000$ is not small and is not sufficiently large.  
   %
   In panel (b), the horizontal axis is the number of measurement trials, $n$.
   Plots in the panel express the scaling of the expected $\delta_{1} $ and $\delta_{2}$ with respect to $n$ with a fixed number of atoms, $N=100$.
   Plots for the separable and GHZ states have a same behavior, i.e., $\Expectation [\delta_{1}] < \Expectation [\delta_{2}]$ for small $n$ and $\Expectation [\delta_{1}] > \Expectation [\delta_{2}]$ for large $n$.
   For small $n$, a correction term, $\frac{8(b-a)}{3(n-1)}\ln \frac{4}{\epsilon}$, in $\delta_{2}$ is not negligible, and $\delta_{1} < \delta_{2}$ holds.
   For large $n$, the correction term becomes negligible, and $\delta_{1} > \delta_{2}$ becomes true because $V_{n} \le V_{\max}$ holds.

%

\end{document}